\documentclass{amsart}
\usepackage{graphicx}
\usepackage{amscd}

\theoremstyle{plain}

\numberwithin{equation}{section}

\input{tcilatex}

\begin{document}
\title[]{Schwarzschild solution on a space-time with torsion}
\author{Gheorghe Zet}
\address{Technical University ''Gh.Asachi'', Department of Physics, Iasi, Romania}
\email{gzet@phys.tuiasi.ro}
\urladdr{}

\begin{abstract}
We obtain the Schwarzschild solution based on teleparallel gravity (TG)
theory formulated in a space-time with torsion only. The starting point is
the Poincar\'{e} gauge theory (PGT).The general structure of TG and its
connection with general relativity (GR) are presented and the Schwarzschild
solution is obtained by solving the field equations of TG. Most of
calculations are performed using the GRTensorII package, running on the
MapleV platform.
\end{abstract}

\maketitle

\section{ Introduction}

In general relativity (GR), the curvature of the space-time is used to
describe the gravitation.The geometry of the space-time replaces the concept
of force. On the other hand, teleparallel gravity (TG) attributes
gravitation to torsion [1,2] and it is a gauge theory for the group of
space-time translations. The gravitational interactions are described in TG
by forces, similar to the Lorentz forces in electrodynamics. Therefore, the
gravitational interactions can be described alternatively in terms of
curvature as is usually done in GR, or in terms of torsion as in TG. It is
believed that requesting a curved or a torsioned space-time to describe
gravity is a matter of convention [3].

In this paper we obtain the Schwarzschild solution in the case of TG. The
general structure of TG and its connection with GR are presented and the
Schwarzschild solution is obtained by solving the field equations. Section 2
contains a review of Poincar\'{e} gauge theory (PGT) on a Minkowski
space-time as base manifold. In Section 3 a non-symmetric connection is
constructed starting with gauge fields of PGT. The field equations of TG are
then obtained in the Section 4 under a general form. The Schwarzschild
solution of these equations and the analytical program are presented in
Section 5. A comparison with GR case is also presented. The conclusions and
some remarks\ are presented in Section 6.

\section{Poincar\'{e} gauge theory}

We will denote the generators of Poincar\'{e} group $P$ by $\{P_{a},M_{ab}\}$%
, where $a,b=0,1,2,3$. Here, $P_{a\text{ }}$ are the generators of
space-time translations and $M_{ab}=-M_{ba}$ are the generators of the
Lorentz rotations. Suppose now that $P$ is a gauge group for gravitation
[4]. Correspondingly, we introduce the 1-form potential $A$ with values in
Lie algebra of the Poincar\'{e} group, defined by formula:$\qquad $

\begin{equation}
A=e^{a}P_{a}+\frac{1}{2}\omega ^{ab}M_{ab}  \tag{1}
\end{equation}
where $e^{a}=$ $e_{\mu }^{a}$ $dx^{\mu }$ and $\omega ^{ab}=\omega _{\mu
}^{ab}$ $dx^{\mu }$ are ordinary 1-forms. The 1-form defines a connection on
the space-time $M_{4}$ of our gauge model with $e_{\mu }^{a}$ and $\omega
_{\mu }^{ab}$ as gauge fields. The 2-form of curvature $F$ is given by the
expression 
\begin{equation}
F=dA+\frac{1}{2}\left[ A,A\right]  \tag{2}
\end{equation}
Inserting (1) in (2) and identifying the result with the definition 
\begin{equation}
F=T^{a}P_{a}+\frac{1}{2}R^{ab}M_{ab}  \tag{3}
\end{equation}

we obtain the 2-forms of torsion $T^{a}$ and of curvature $R^{ab}$ in the
form 
\begin{equation}
T^{a}=de^{a}+\omega ^{a}\text{ }_{b}\wedge e^{b},  \tag{4}
\end{equation}
and respectively 
\begin{equation}
R^{a}\text{ }_{b}=d\omega ^{a}\text{ }_{b}+\omega ^{a}\text{ }_{c}\wedge
\omega ^{c}\text{ }_{b}.  \tag{5}
\end{equation}
We use the Minkowski metric $\eta _{ab}=diag\left( 1,-1,-1,-1\right) $ on
the Poincar\'{e} group manifold to rise and lower the latin indices $a,$ $b,$
$c$. Written on the components, the equations (4) and (5) give: 
\begin{equation}
T_{\mu \nu }^{a}=\partial _{\mu }e_{\nu }^{a}-\partial _{\nu }e_{\mu
}^{a}+\left( \omega _{\mu }^{ab}e_{\nu }^{c}-\omega _{\nu }^{ab}e_{\mu
}^{c}\right) \eta _{bc},  \tag{6}
\end{equation}
and respectively 
\begin{equation}
R^{ab}\text{ }_{\mu \nu }=\partial _{\mu }\omega _{\nu }^{ab}-\partial _{\nu
}\omega _{\mu }^{ab}+\left( \omega _{\mu }^{ac}\omega _{\nu }^{db}-\omega
_{\nu }^{ac}\omega _{\mu }^{db}\right) \eta _{cd}.  \tag{7}
\end{equation}
The quantity $T_{\mu \nu }^{a}$ is the torsion tensor and the quantity $%
R^{ab}$ $_{\mu \nu }$ is the curvature tensor of the connection $A$ defined
by the equation (1). The connection $A$ defines a structure Einstein-Cartan
(EC) on the space-time and we will denote the corresponding space by $U_{4}$%
. This space have both torsion and curvature. In the next Section, we will
consider the case of a space-time with non-null torsion and vanishing
curvature to construct the teleparallel theory of gravity ($TG$).

\section{ Teleparallel gravity}

We interpret $e_{\nu }^{a}$ as tetrad fields and $\omega _{\nu }^{ab}$ as
spin connection. A model of gauge theory based on Poincar\'{e} group and
implying only torsion can be obtained choosing $\omega _{\nu }^{ab}=0$. Then
the curvature tensor $R^{ab}$ $_{\mu \nu }$ vanishes and the torsion in
equation (6) becomes: 
\begin{equation}
T^{a}\text{ }_{\mu \nu }=\partial _{\mu }e_{\nu }^{a}-\partial _{\nu }e_{\mu
}^{a}  \tag{8}
\end{equation}
Expressed in a coordinate basis, this tensor has the components: 
\begin{equation}
T^{\rho }\text{ }_{\mu \nu }=e_{a}\text{ }^{\rho }\partial _{\mu }e_{\nu
}^{a}-e_{a}\text{ }^{\rho }\partial _{\nu }e_{\mu }^{a},  \tag{9}
\end{equation}
where $e_{a}$ $^{\rho }$ is the inverse of $e^{a}$ $_{\rho }$.

Now, we define the Cartan connection $\Gamma $ on the space-time $M_{4}$
with nonsymmetric coefficients 
\begin{equation}
\Gamma ^{\rho }\text{ }_{\mu \nu }=e_{a}\text{ }^{\rho }\partial _{\nu
}e_{\mu }^{a}.  \tag{10}
\end{equation}
This definition is suggested by the expression (9) of the torsion
components. Therefore, the connection $\Gamma $ has the torsion given by the
usually formula 
\begin{equation}
T^{\rho }\text{ }_{\mu \nu }=\Gamma ^{\rho }\text{ }_{\nu \mu }-\Gamma
^{\rho }\text{ }_{\mu \nu }.  \tag{11}
\end{equation}
With respect to the connection $\Gamma $, the tetrad field is parallel, that
is: 
\begin{equation}
\nabla _{\mu }e^{a}\text{ }_{\nu }=\partial _{\mu }e_{\nu }^{a}-\Gamma
^{\rho }\text{ }_{\nu \mu }\text{ }e^{a}\text{ }_{\rho }=0.  \tag{12}
\end{equation}

Curvature and torsion have to be considered as properties of the connections
and therefore many different connections are allowed on the same space-time
[5]. For example, starting with the tetrad field $e^{a}$ $_{\mu }$ we can
define the Riemannian metric: 
\begin{equation}
g_{\mu \nu }=\eta _{ab}\text{ }e^{a}\text{ }_{\mu }\text{ }e^{a}\text{ }%
_{\nu }.  \tag{13}
\end{equation}
Then, we can introduce the Levi-Civita connection 
\begin{equation}
\overset{\circ }{\Gamma }_{\mu \nu }^{\sigma }=\frac{1}{2}g^{\sigma \rho
}\left( \partial _{\mu }g_{\rho \nu }+\partial _{\nu }g_{\rho \mu }-\partial
_{\rho }g_{\mu \nu }\right) .  \tag{14}
\end{equation}
This connection is metric preserving: 
\begin{equation}
\overset{\circ }{\nabla _{\rho }}g^{\mu \nu }=\partial _{\rho }g+\overset{%
\circ }{\Gamma }_{\sigma \rho }^{\mu }g^{\sigma \nu }+\overset{\circ }{%
\Gamma }_{\sigma \rho }^{\nu }g^{\sigma \mu }=0.  \tag{15}
\end{equation}
The relation between the two connections $\Gamma $ and $\overset{\circ }{%
\Gamma }$ is 
\begin{equation}
\Gamma ^{\sigma }\text{ }_{\mu \nu }=\overset{\circ }{\Gamma }_{\mu \nu
}^{\sigma }+K^{\sigma }\text{ }_{\mu \nu },  \tag{16}
\end{equation}
where 
\begin{equation}
K^{\sigma }\text{ }_{\mu \nu }=\frac{1}{2}\left( T_{\mu }\text{ }^{\sigma }%
\text{ }_{\nu }+T_{\nu }\text{ }^{\sigma }\text{ }_{\mu }-T^{\sigma }\text{ }%
_{\mu \nu }\right)  \tag{17}
\end{equation}
is the contortion tensor.

The curvature tensor of the Levi-Civita Connection $\overset{\circ }{\Gamma }
$ is: 
\begin{equation}
\overset{\circ }{R}^{\sigma }\text{ }_{\rho \mu \nu }=\partial _{\mu }%
\overset{\circ }{\Gamma }_{\rho \nu }^{\sigma }+\overset{\circ }{\Gamma }%
_{\tau \mu }^{\sigma }\overset{\circ }{\Gamma }_{\rho \nu }^{\tau }-\left(
\mu \leftrightarrow \nu \right)  \tag{18}
\end{equation}
Because the connection coefficients $\overset{\circ }{\Gamma }_{\mu \nu
}^{\sigma }$are symmetric in the indices $\mu $ and $\nu $, its torsion is
vanishing.Therefore, the Levi-Civita connection $\overset{\circ }{\Gamma }$
have non-null curvature, but no torsion. Contrarily, the Cartan connection $%
\Gamma $ presents torsion, but no curvature. Indeed, using the definition
(10), we can verify that the curvature of the connection $\Gamma $ vanishes
identically: 
\begin{equation}
\overset{}{R}^{\sigma }\text{ }_{\rho \mu \nu }=\partial _{\mu }\overset{}{%
\Gamma }_{\rho \nu }^{\sigma }+\overset{}{\Gamma }_{\tau \mu }^{\sigma }%
\overset{}{\Gamma }_{\rho \nu }^{\tau }-\left( \mu \leftrightarrow \nu
\right) \equiv 0.  \tag{19}
\end{equation}
Then, substituting (16) into the expression (19), we obtain: 
\begin{equation}
R^{\sigma }\text{ }_{\rho \mu \nu }=\overset{\circ }{R}^{\sigma }\text{ }%
_{\rho \mu \nu }+Q^{\sigma }\text{ }_{\rho \mu \nu }\equiv 0,  \tag{20}
\end{equation}
where 
\begin{equation}
Q^{\sigma }\text{ }_{\rho \mu \nu }=D_{\mu }K^{\sigma }\text{ }_{\rho \nu
}+\Gamma ^{\sigma }\text{ }_{\tau \nu }K^{\tau }\text{ }_{\rho \mu }-\left(
\mu \leftrightarrow \nu \right)  \tag{21}
\end{equation}
is the non-metricity tensor. Here 
\begin{equation}
D_{\mu }K^{\sigma }\text{ }_{\rho \nu }=\partial _{\mu }K^{\sigma }\text{ }%
_{\rho \nu }+\Gamma ^{\sigma }\text{ }_{\tau \mu }K^{\tau }\text{ }_{\rho
\nu }-\Gamma ^{\sigma }\text{ }_{\tau \nu }K^{\tau }\text{ }_{\rho \mu } 
\tag{22}
\end{equation}
is the teleparallel covariant derivative.

The equation (20) has an interesting interpretation [6]: the contribution $%
\overset{\circ }{R}^{\sigma }$ $_{\rho \mu \nu }$ coming from the
Levi-Civita connection $\overset{\circ }{\Gamma }$ compensates exactly the
contribution $Q^{\sigma }$ $_{\rho \mu \nu }$ coming from the Cartan
connection $\Gamma $, yielding an identically zero Cartan curvature tensor $%
R^{\sigma }$ $_{\rho \mu \nu }$.

Now, according to GR theory, the dynamics of the gravitational field is
determined by the Lagrangian [6]: 
\begin{equation}
L_{GR}=\frac{\sqrt{-g}c^{4}}{16\pi G}\overset{\circ }{R},  \tag{23}
\end{equation}
where $\overset{\circ }{R}$ $=$ $g^{\mu \nu }\overset{\circ }{R}^{\rho }$ $%
_{\mu \rho \nu }$ is the scalar curvature of the Levi-Civita connection $%
\overset{\circ }{\Gamma }$, $G$ is the gravitational constant and $g=\det
\left( g_{\mu \nu }\right) $. Then, substituting $\overset{\circ }{R}$ as
obtained from (20), one obtains up to divergences [6]: 
\begin{equation}
L_{TG}=\frac{ec^{4}}{16\pi G}S^{\rho \mu \nu }T_{\rho \mu \nu }\text{ }, 
\tag{24}
\end{equation}
where $e=\det (e^{a}$ $_{\mu })=\sqrt{-g}$, and 
\begin{equation}
S^{\rho \mu \nu }=-S^{\rho \nu \mu }=\frac{1}{2}\left( K^{\mu \nu \rho
}-g^{\rho \nu }T^{\sigma \mu }\text{ }_{\sigma }+g^{\rho \mu }T^{\sigma \nu }%
\text{ }_{\sigma }\right)  \tag{25}
\end{equation}
is a tensor written in terms of the Cartan connection only. The equation
(24) gives the Lagrangian of the TG as a gauge theory of gravitation for the
translation group.

It is proven [7] that the translational gauge theory of gravitation TG with
the Lagrangian $L_{TG}$ quadratic in torsion is completely equivalent to
general relativity GR with usual Lagrangian $L_{GR}$ linear in the scalar
curvature. Therefore, the gravitation presents two equivalent descriptions:
one GR in terms of a metric geometry and another one TG in which the
underlying geometry is provided by a teleparallel structure.

In the next Section we will obtain the field equations of gravitation within
TG theory.

\section{Field equations}

Taking the variation of the Lagrangian $L_{TG}$ in Eq. (24) with respect to
the gauge field $e^{a}$ $_{\mu }$, one obtains the teleparallel version of
the gravitational field equations: 
\begin{equation}
\partial _{\nu }\left( eS_{a}\text{ }^{\nu \rho }\right) -\frac{4\pi G}{c^{4}%
}\left( ej_{a}\text{ }^{\rho }\right) =0,  \tag{26}
\end{equation}
where $S_{a}$ $^{\nu \rho }=e_{a}$ $^{\mu }S_{\mu }$ $^{\nu \rho }$ and 
\begin{equation*}
S_{\mu }\text{ }^{\nu \rho }=g_{\mu \tau }S^{\tau \nu \rho }=\frac{1}{4}%
\left( T_{\mu }\text{ }^{\nu \rho }+T^{\nu }\text{ }_{\mu }\text{ }^{\rho
}-T^{\rho }\text{ }_{\mu }\text{ }^{\nu }\right) -\frac{1}{2}\left( \delta
_{\mu }\text{ }^{\rho }T_{\sigma }\text{ }^{\nu \sigma }-\delta _{\mu }\text{
}^{\nu }T_{\sigma }\text{ }^{\rho \sigma }\right) .
\end{equation*}

The quantity $j_{a}$ $^{\rho }$ in Eq.(26) is the gauge gravitational
current, defined analogous to the Yang-Mills theory: 
\begin{equation}
j_{a}\text{ }^{\rho }=\frac{1}{e}\frac{\partial L_{TG}}{\partial e_{a}\text{ 
}^{\rho }}=-\frac{c^{4}}{4\pi G}e_{a}\text{ }^{\sigma }S_{\mu }\text{ }^{\nu
\rho }T^{\mu }\text{ }_{\nu \sigma }+\frac{1}{e}e_{a}\text{ }^{\rho }L_{TG.}
\tag{27}
\end{equation}
The current $j_{a}$ $^{\rho }$ represents the energy-momentum of the
gravitational field. The term $eS_{a}$ $^{\sigma \rho }$ is called
superpotential in the sense that its derivative yields the gauge current $%
ej_{a}$ $^{\rho }$ . Due to the anti-symmetry of $S_{a}$ $^{\sigma \rho }$
in the indices $\sigma $ and $\rho $ the quantity $ej_{a}$ $^{\rho }$ is
conserved as a consequence of the field equations, i.e. 
\begin{equation}
\partial _{\rho }\left( ej_{a}^{\text{ \ }\rho }\right) =0.  \tag{28}
\end{equation}
Making use of Eq. (10) to express $\partial _{\rho }e_{a}$ $^{\sigma }$, the
field equations (26) can be written in a purely space-time form: 
\begin{equation}
\frac{1}{e}\partial _{\sigma }\left( eS_{\mu }\text{ }^{\sigma \rho }\right)
-\frac{4\pi G}{c^{4}}\left( t_{\mu }\text{ }^{\rho }\right) =0,  \tag{29}
\end{equation}
where $t_{\mu }$ $^{\rho }$ is the canonical energy-momentum pseudo-tensor
of the gravitational field [5], defined by the expression: 
\begin{equation}
t_{\sigma }^{\text{ \ \ }\rho }=\frac{c^{4}}{4\pi G}\Gamma ^{\mu }\text{ }%
_{\nu \sigma }S_{\mu }\text{ }^{\nu \rho }+\frac{1}{e}\delta _{\sigma }^{%
\text{ \ }\rho }L_{TG}.  \tag{30}
\end{equation}
It is important to notice that the canonical energy-momentum pseudo-tensor $%
t_{\mu }$ $^{\rho }$ is not simply the gauge current $j_{a}^{\text{ \ }\rho
} $ with the Lorentz index ''$a$'' changed to the space-time index ''$\mu $%
''. It incorporates also an extra term coming from the derivative term of
Eq. (26) 
\begin{equation}
t_{\sigma }^{\text{ \ \ }\rho }=e_{\text{ \ }\sigma }^{a}j_{a}^{\text{ \ \ }%
\rho }+\frac{c^{4}}{4\pi G}\Gamma ^{\mu }\text{ }_{\sigma \nu }S_{\mu }\text{
}^{\nu \rho }.  \tag{31}
\end{equation}
Like the gauge current $ej_{a}^{\text{ \ }\rho }$, the pseudo-tensor $%
et_{\mu }$ $^{\rho }$ is conserved as a consequence of the field equation: 
\begin{equation}
\partial _{\rho }\left( et_{\mu }^{\text{ \ \ }\rho }\right) =0.  \tag{32}
\end{equation}
But, due to the pseudo-tensor character of $t_{\mu }$ $^{\rho }$, this
conservation law can not be expressed with a covariant derivative, in
contrast with $j_{a}^{\text{ \ }\rho }$ case.

Using the previous results, we will prove in the next Section that the
Schwarzschild solution can be obtained from the field equations (29) of the
teleparallel theory of gravity.

\section{Schwarzschild solution and analytical program}

Because we are looking for a spherically symmetric solution of the field
equations, we will choose the Minkowski metric 
\begin{equation}
ds^{2}=dt^{2}-dr^{2}-r^{2}\left( d\theta ^{2}+\sin ^{2}\theta d\varphi
^{2}\right)  \tag{33}
\end{equation}
on the space-time manifold. The coordinates $x^{0},x^{1},x^{2},x^{3}$
correspond to $ct,r,\theta ,\varphi $ respectively. Then we will consider
the gauge theory based on Poincar\'{e} group with $\omega _{\nu }^{ab}=0$
described in Section 3. The tetrad field $e_{\text{ \ }\mu }^{a}$ will be
chosen under the form: 
\begin{equation}
(e_{\text{ \ }\mu }^{a})=\left( 
\begin{array}{cccc}
e^{A/2} & 0 & 0 & 0 \\ 
0 & e^{B/2} & 0 & 0 \\ 
0 & 0 & r & 0 \\ 
0 & 0 & 0 & r\sin \theta
\end{array}
\right) ,  \tag{34}
\end{equation}
where $A=A(r)$ and $B=B(r)$ are functions only of the 3D radius $r$. The
inverse of $e_{\text{ \ }\mu }^{a}$ is therefore: 
\begin{equation}
(e_{a}^{\text{ \ \ }\mu })=\left( 
\begin{array}{cccc}
e^{-A/2} & 0 & 0 & 0 \\ 
0 & e^{-B/2} & 0 & 0 \\ 
0 & 0 & \frac{1}{r} & 0 \\ 
0 & 0 & 0 & \frac{1}{r\sin \theta }
\end{array}
\right)  \tag{35}
\end{equation}
The metric $g_{\mu \nu }$ $=\eta _{ab}$ $e^{a}$ $_{\mu }$ $e^{a}$ $_{\nu }$
will have then the form: 
\begin{equation}
\left( g_{\mu \nu }\right) =\left( 
\begin{array}{cccc}
e^{A} & 0 & 0 & 0 \\ 
0 & -e^{B} & 0 & 0 \\ 
0 & 0 & -r^{2} & 0 \\ 
0 & 0 & 0 & -r^{2}\sin ^{2}\theta
\end{array}
\right)  \tag{36}
\end{equation}
where the Eq. (33) have been used. The inverse of $g_{\mu \nu }$ is
evidently: 
\begin{equation}
\left( g^{\mu \nu }\right) =\left( 
\begin{array}{cccc}
e^{-A} & 0 & 0 & 0 \\ 
0 & -e^{-B} & 0 & 0 \\ 
0 & 0 & -\frac{1}{r^{2}} & 0 \\ 
0 & 0 & 0 & -\frac{1}{r^{2}\sin ^{2}\theta }
\end{array}
\right) .  \tag{37}
\end{equation}

We use the above expressions to compute the coefficients $\Gamma ^{\rho }$ $%
_{\mu \nu }$ of the Cartan connection, the components $T_{\mu }^{\text{ \ \ }%
\nu \rho }$ of the torsion tensor, of the tensor $S_{\mu }^{\text{ \ \ }\nu
\rho }$ and of the canonical energy-momentum pseudo-tensor $t_{\mu }$ $%
^{\rho }$. From this point at end we performed all the calculations using an
analytical program conceived by us and which is given in Section 6. For
example, the non-null components of the tensor $T_{\mu }^{\text{ \ \ }\nu
\rho }$ are: 
\begin{equation}
T_{0}^{\text{ \ \ }01}=\frac{A^{\prime }e^{-B}}{2},\text{ \ \ \ \ }T_{2}^{%
\text{ \ \ }21}=T_{3}^{\text{ \ \ }31}=\frac{e^{-B}}{r},\text{ \ \ }T_{3}^{%
\text{ \ \ }32}=\frac{\cot \theta }{r^{2}},  \tag{38}
\end{equation}
where $A^{\prime }=\frac{dA}{dr}$ denote de derivative of the function $A(r)$
with respect to the variable $r$. We will use the same notation for the
derivative of $B(r)$, that is $B^{\prime }=\frac{dB}{dr}.$

We list also the non-null components of the tensor $S_{\mu }$ $^{\sigma \rho
}$ and of the canonical energy-momentum pseudo-tensor $t_{\mu }$ $^{\rho }$
of the gravitational field. Thus, for $S_{\mu }$ $^{\sigma \rho }$ we have: 
\begin{equation*}
S_{0}\text{ }^{10}=\frac{e^{-B}}{r},\text{ \ }S_{0}\text{ }^{20}=S_{1}\text{ 
}^{21}=\frac{\cot \theta }{2r^{2}},\text{ \ }S_{2}\text{ }^{12}=S_{3}\text{ }%
^{13}=\frac{e^{-B}\left( rA^{\prime }+2\right) }{4r},
\end{equation*}
and for $t_{\mu }$ $^{\rho }$: 
\begin{eqnarray*}
t_{0}\text{ }^{0} &=&t_{1}\text{ }^{1}=t_{2}\text{ }^{2}=t_{3}\text{ }^{3}=%
\frac{c^{4}}{4\pi G}\frac{e^{-B}\left( rA^{\prime }+1\right) }{2r^{2}},\text{
} \\
\text{\ }t_{1}\text{ }^{2} &=&-\frac{c^{4}}{4\pi G}\frac{\left( A^{\prime
}+B^{\prime }\right) \cot \theta }{4r^{2}},\text{ }t_{2}\text{ }^{1}=-\frac{%
c^{4}}{4\pi G}\frac{e^{-B}\left( rA^{\prime }+2\right) \cot \theta }{4r}.
\end{eqnarray*}

Now, using these components we obtain from (29) the following equations of
gravitational field in $TG$ theory: 
\begin{equation}
e^{-B}\left( \frac{B^{\prime }}{r}-\frac{1}{r^{2}}\right) +\frac{1}{r^{2}}=0,
\tag{39}
\end{equation}
\begin{equation}
e^{-B}\left( \frac{A^{\prime }}{r}+\frac{1}{r^{2}}\right) -\frac{1}{r^{2}}=0,
\tag{40}
\end{equation}
\begin{equation}
2A^{\prime \prime }+(\frac{2}{r}+A^{\prime })(A^{\prime }-B^{\prime })=0, 
\tag{41}
\end{equation}
where $A^{\prime \prime }=\frac{d^{2}A}{dr^{2}}$ is the second derivative of 
$A(r)$ with respect to $r$. It is easy to verify that the third field
equation (41) is a combination of the first two (39) and (40). Therefore,
the equation (39) and (40) are the only independent field equations and they
determine the two unknown functions $A(r)$ and $B(r)$.

The solution of the equations (39) and (40) are [8]: 
\begin{equation}
e^{-B}=e^{A}=1+\frac{\alpha }{r},  \tag{42}
\end{equation}
where $\alpha $ is a constant of integration. It is known that the constant $%
\alpha $ can be expressed by the mass $m$ of the body which is the source of
the gravitational field with spherically symmetry [8]: $\alpha =-\frac{2Gm}{%
c^{2}}$. Therefore, we obtain the Scwarzschild solution 
\begin{equation}
ds^{2}=(c^{2}-\frac{2Gm}{r})dt^{2}-\frac{dr^{2}}{\left( 1-\frac{2Gm}{c^{2}r}%
\right) }-r^{2}\left( d\theta ^{2}+\sin ^{2}\theta d\varphi ^{2}\right) , 
\tag{43}
\end{equation}
within the frame-work of $TG$ theory of the gravitational field.

Finally, we emphasis again on the conclusion given in Ref. [2]: the
gravitation presents two equivalent descriptions, one in terms of a metric
geometry, and another one in which the underlying geometry is that provided
by a teleparallel structure.

\bigskip Program ''TELEPARALLEL GRAVITY.MWS''

restart: grtw( ):

grload (minkowski, `spheric.mpl');

grdef(`ev\{\symbol{94}a miu\}'); grcalc(ev(up,dn));

grdef(`evinv\{ a \symbol{94}miu\}'); grcalc(evinv(dn,up));

grdef(`eta\{a b\}`); grcalc(eta(dn,dn));

grdef(`etainv\{\symbol{94}a \symbol{94}b\}`); grcalc(etainv(up,up));

grdef(`ge\{miu niu\}:=ev\{\symbol{94}a miu\}*ev\{\symbol{94}betha
niu\}*eta\{a b\}`); grcalc(ge(dn,dn));

grdef(`geinv\{\symbol{94}miu \symbol{94}niu\}:=etainv\{\symbol{94}a \symbol{%
94}b\}*evinv\{a \symbol{94}miu\}*evinv\{b \symbol{94}niu\}`);
grcalc(geinv(up,up));

grdef(`gama\{\symbol{94}sigma miu niu\} :=evinv\{a \symbol{94}sigma\}*ev\{%
\symbol{94}a miu,niu\}`);

grcalc(gama(up,dn,dn));

grdef(`TORS\{\symbol{94}sigma miu niu\}:=gama\{\symbol{94}sigma niu
miu\}-gama\{\symbol{94}sigma miu niu\}`);

grcalc(TORS(up,dn,dn));

grdef(`TS1\{miu \symbol{94}niu \symbol{94}rho\}:=ge\{miu sigma\}*geinv\{%
\symbol{94}niu \symbol{94}lambda\}*geinv\{\symbol{94}rho \symbol{94}tau\}*

TORS\{\symbol{94}sigma lambda tau\}`); grcalc(TS1(dn,up,up));

grdef(`TS2\{\symbol{94}niu miu \symbol{94}rho\}:=geinv\{\symbol{94}rho 
\symbol{94}sigma\}*TORS\{\symbol{94}niu miu sigma\}`); grcalc(TS2(up,dn,up));

grdef(`S\{miu \symbol{94}niu \symbol{94}rho\}:=(1/4)*(TS1\{miu \symbol{94}%
niu \symbol{94}rho\}+TS2\{\symbol{94}niu miu \symbol{94}rho\}-TS2\{\symbol{94%
}rho miu

\symbol{94}niu\})-(1/2)*(kdelta\{miu \symbol{94}rho\}*TS1\{sigma \symbol{94}%
niu \symbol{94}sigma\}-kdelta\{miu \symbol{94}niu\}*TS1\{sigma \symbol{94}%
rho \symbol{94}sigma\})`);

grcalc(S(dn,up,up));

grdef(`ed:=r\symbol{94}2*sin(theta)*exp((A(r)+B(r))/2); grcalc(e);

grdef(`eS\{lambda \symbol{94}sigma \symbol{94}rho\}:=ed*S\{lambda \symbol{94}%
sigma \symbol{94}rho\}`);

grcalc(eS(dn,up,up));

grdef(`t\{lambda \symbol{94}rho\}:=(c\symbol{94}4/(4*pi*G))*(gama\{\symbol{94%
}miu niu lambda\}*S\{miu \symbol{94}niu \symbol{94}rho\}+(1/4)*kdelta\{lambda

\symbol{94}rho\}*S\{tau \symbol{94}miu \symbol{94}niu\}*TORS\{\symbol{94}tau
miu niu\})`); grcalc(t(dn,up));

grdef(`EQ\{lambda \symbol{94}rho\}:=(1/ed)*eS\{lambda \symbol{94}sigma 
\symbol{94}rho,sigma\}-(4*pi*G/c\symbol{94}4)*t\{lambda \symbol{94}rho\}`);

grcalc(EQ(dn,up)); grdisplay(\_);

\section{Concluding remarks}

We obtained the Schwarzschild solution within the teleparallel theory\ (TG)
of gravity which is formulated in a space-time with torsion only. This can
be interpreted as an indication that source of the torsion can be also the
mass of the bodies that create the gravitational field, not only the spin.
Therefore the torsion and curvature of the space-time is determined by the
mater distribution in the considered region.

Most of the calculations have been performed using an analytical program
conceived by us. The program allows to calculate the components of all
quantities appearing in the model and to obtain also the equations of the
gravitational field.

The TG theory can be used also to unify the gravitational field with other
fundamental interactions (electromagnetic, weak and strong). Some results
about this problem are given in our paper [9].

\bigskip

\section{References}

1. Calcada, M., Pereira, J.G.: Int.J.Theor.Phys.\ Vol. 41, p.729, 2002

2. Andrade, V.C., Pereira, J.G.: Torsion and electromagnetic field,
arXiv:qr-qc/9708051, v2-11 Jan. 1999

3. Capozziello, S., Lambiase, G., Stornaiolo: Ann. Phys. (Leipzig) Vol. 10,
p.8, 2001

4. Zet, G., Manta, V.: Int.J.Modern Phys. C, Vol.13, p. 509, 2002

5. Blagojevic, M.: Three lectures on Poincar\'{e} gauge theory,
arXiv:qr-qc/0302040, v1-11 Feb. 2003

6. Andrade, V.C., Guillen, L.C.T., Pereira, J.G.: Teleparallel gravity:an
overview, arXiv:qr-qc/0011087, 2000

7. Andrade, V.C., Pereira, J.G.: Phys.Rev. D56, p. 4689, 1997

8. Landau, L., Lifchitz, E.: Th\'{e}orie du champ, Ed. Mir, Moscou, 1966

9. Zet, G.: Unified theory of fundamental interactions in a space-time with
torsion (in preparation)

\end{document}